\documentclass[aps,pra,twocolumn,showpacs,superscriptaddress,groupedaddress]{revtex4}
\usepackage{graphicx} 
\usepackage{amsmath}

\begin{document}

\title{Quantum droplet speed management and supersolid behavior in external harmonic confinement}
\author{Saurab Das}
\author{Ajay Nath}
\affiliation{Indian Institute of Information Technology Vadodara, Gujarat, India 382 028}

\begin{abstract}
In this work, we propose a management method for controlling the speed and direction of self-bound quantum droplets (QDs) in a binary Bose-Einstein condensate mixture under time-modulated external harmonic confinement. Utilizing the 1D extended Gross-Pit"{a}evskii equation, QDs are constructed within both regular and expulsive parabolic traps, considering temporally varying attractive quadratic beyond mean field and repulsive cubic mean-field atom-atom interactions. Through the derived wavefunction solution, we illustrate the dynamics of slowing, stopping, reversing, fragmentation, collapse, and revival of droplets. Additionally, the solutions reveal a crystalline order with a superfluid background, indicative of supersolid behavior in various parameter domains. Notably, one-third of the constant background matches the lowest residual condensate. These findings hold potential applications in matter-wave interferometry and quantum information processing. 

\end{abstract}

\maketitle

\section{Introduction}

In recent years, quantum droplets (QDs), which are the quantum analogs of classical fluid droplets arising due to quantum fluctuations, have garnered significant attention in the literature of Bose-Einstein condensates (BEC) and superfluids \cite{Luo,ttcher,Malomed1,Khan1,Rakshit}. QDs are unique in being self-bound and exhibiting liquid-like properties such as incompressibility and surface tension, making them the only atomic species that naturally maintain liquidity at zero temperature. Following the theoretical proposal by Petrov, QDs have been experimentally realized in dipolar BECs (Er and Dy) with anisotropic interactions between dipolar atoms \cite{Kadau,Schmitt,Schmitt1}, as well as in homonuclear ($^{39}$K) and heteronuclear ($^{41}$K-$^{87}$Rb, $^{23}$Na-$^{87}$Rb) two-component BECs with isotropic contact interactions \cite{Cabrera,Cheiney,Semeghini}. The self-binding of droplets in free space is predicted to occur through a delicate balance between the vanishing effective mean field (EMF) from attractive interatomic interactions and repulsive beyond mean field (BMF) effects due to quantum fluctuations, which become increasingly significant as density increases \cite{Lee,Petrov,Petrov1}. The experimental realization of QDs has led to explorations across diverse contexts within BECs: droplet dynamics in one-dimensional (1D) systems \cite{Astrakharchik}, pair superfluid droplets in 1D optical lattices \cite{Ivan,Ivan1}, droplet-to-soliton crossover \cite{Maitri2}, dimensional crossover in droplets \cite{Zin,Lavoine}, collective excitations of 1D QDs \cite{Tylutki}, spherical droplets \cite{Hu}, dark QDs \cite{Edmonds}, vortex QDs \cite{Zhang}, droplet scattering \cite{Xia}, quantum droplet molecules \cite{Elhad}, and supersolidity in 1D QDs \cite{ttcher,Parit}. However, less attention has been directed towards the theoretical control of QD wave speed in the presence of various external confinements.

The dynamics of nonlinear excitations in BECs are theoretically studied using the three-dimensional Gross-Pitaevskii equation (GPE) under the mean-field approximation. Numerous investigations have been reported on managing nonlinearity and external traps \cite{Usama,Dalfovo,Pethick}. Early studies focused on the one-dimensional (1D) GPE with cubic or cubic-quintic nonlinearities for high-density condensate atoms \cite{Atre,Beitia,Sulem,Rajaraman,Beitia1,Wang,Ajay,Roy,Tang,Baines}, with significant contributions from Serkin et al. on solitary solutions using inverse scattering \cite{Serkin}, and from Kruglov et al. on self-similar solutions in nonlinear fibers \cite{Kruglov}. However, recent observations of QDs necessitate modifications to the GPE to accurately describe their stabilization mechanism through BMF contributions. The BMF effect is represented by the three-dimensional extended Gross-Pitaevskii equation (eGPE) model, which incorporates first-order Lee-Huang-Yang corrections \cite{Bisset,Petrov,Petrov1}. The eGPE used to model dipolar condensates aligns well with ab initio quantum Monte Carlo calculations for both dipolar droplets and those formed in binary BECs dominated by contact interactions \cite{Cinti,Macia,Boronat}.

Although QDs are self-bound, in practice, their formation within a BEC requires a confining trap. When the potential strongly confines atoms in two directions, BEC atoms can be effectively restricted to a single dimension. Utilizing one-dimensional geometry allows for a broader range of interaction strength manipulation compared to the three-dimensional scenario, primarily due to the suppression of three-body losses. Consequently, quantum effects are amplified in the 1D system, enhancing stability. Unlike in 3D scenarios, 1D quantum droplets emerge in a regime where the mean-field interaction is generally repulsive, and quantum fluctuations induce an effective quadratic nonlinear attraction term that can be represented by the 1D eGPE \cite{Petrov,Tylutki}. The 1D eGPE has been used to investigate collective excitations \cite{Tylutki}, droplet-to-soliton transitions in harmonic traps \cite{Maitri,Bhatia}, enhanced mobility \cite{Kartashov}, the formation of periodic matter wave droplet lattices in optical lattices \cite{Maitri1}, the generation of kink solitons \cite{Shukla}, modulational instability \cite{Mithun1}, and various other contexts \cite{Malomed1,Khan1}.

Motivated by the generation of slow light and soliton management techniques in ultracold atoms \cite{Hau,Dutton,Kengne}, this article investigates the possibility of slowing, stopping, reversing, fragmentation, and collapse and revival of quantum droplets (QDs) in a two-component Bose-Einstein condensate (BEC) mixture. We consider QDs generated in a two-component BEC under external harmonic confinement using the 1D eGPE. We calculate the exact analytical forms of the wavefunction, phase, and effective mean field (EMF)/beyond mean field (BMF) nonlinearities. Interestingly, we map the consistency condition governing the droplet profile to the linear Schrödinger eigenvalue problem and the Riccati equation. This approach allows us to analytically solve the 1D eGPE for a wide range of temporal variations in control parameters. By appropriately tuning the oscillator frequency of the chosen trap, we illustrate the slowing, stopping, reversing, fragmentation, and collapse/revival of QDs. Furthermore, the solutions reveal a crystalline order with a superfluid background, indicative of supersolid behavior across different parameter domains. Notably, one-third of the constant background corresponds precisely to the lowest residual condensate. Here, the QDs form a periodic array with a constant inter-droplet spacing, immersed in a residual BEC background, which establishes inter-droplet coherence.

In this paper, we investigate the application of an external harmonic trap to alter the velocity of QDs. This technique demonstrates the ability to slow, stop, or even reverse the direction of QD wave propagation. Section II outlines the general methodology for managing droplet wave speeds, while Section III applies this approach to manipulate the velocities of QDs. In Section IV, we explore the supersolid behavior of droplets under scenarios of slowing, stopping, reversing, and collapsing/reviving. The stability of the obtained solutions is analyzed in Section V. Finally, concluding remarks are provided in Section VI.

\begin{figure*}[htpb]
\centering
\includegraphics[scale=0.94]{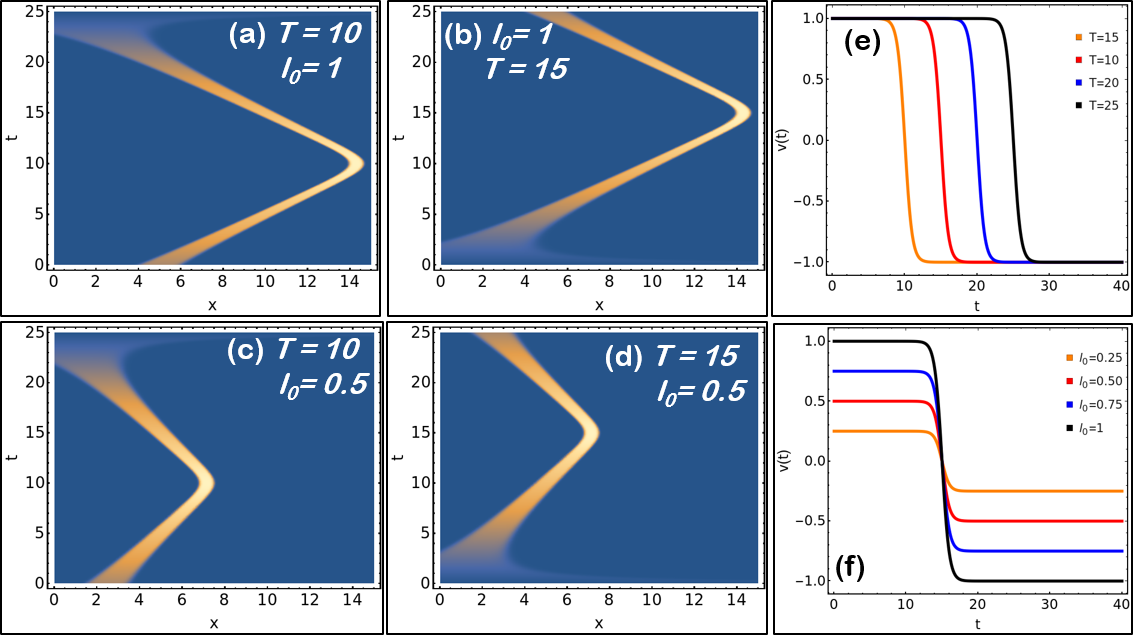}
\caption{\label{fig2} Reversing of QDs in parabolic confinement. Here, the condensate density is plotted with changing magnitude of oscillator frequency ($T$ and $l_{0}$) for (a) $T=10$, and $l_{0}=1$, (b) $T=15$, and $l_{0}=1$, (c) $T=10$, and $l_{0}=0.5$, and (d) $T=15$, and $l_{0}=0.5$. The corresponding velocity with respect to time for varying magnitude of (e) $T$ with constant $l_{0}=1$, and (f) $l_{0}$ with constant $T=15$ are illustrated. The position and time of droplet reversing can be tuned by oscillator frequency. For each case, $A_{0}=4$, $\lambda_{0}=-2/9$, $\lambda_{1}=-2/9$, $\lambda_{2}=-1$, $\lambda_{3}=0.99999999$ and the spatial coordinate is scaled by the oscillator length.}
\end{figure*}

\section{Analytical model and framework}
We begin by considering a 1D homonuclear binary BEC mixture confined under a time-modulated external harmonic trap $V(x,t)=(1/2) M(t) x^{2}$ with temporally varying oscillator frequency $M(t)$. Depending on $M(t) > 0$ or $M(t) < 0$, the external trap is confining or expulsive. The mixture consists of two distinct hyperfine states of $^{39}$K \cite{Semeghini}, confined tightly in the transverse directions while the {\it x}-direction remains elongated and unconfined. To simplify, both states of the binary BEC are mutually symmetric, represented by $\psi_{1}=\psi_{2}=c_{0} \psi$, with equal atom numbers $N_{1}$=$N_{2}$=$N$, and equal masses. Additionally, we assume equal intra-atomic coupling constants $g_{11} = g_{22}\equiv g (>0)$, given by $2 \hbar^{2} a_{s}(x, t)/(m a^{2}_{\perp})$, and an inter-atomic coupling constant $ g_{12} (<0)$, contributing to a droplet environment where $ \delta g = g_{12} + g > 0$  \cite{Astrakharchik}. These assumptions allow the binary BEC mixture to be described by a simplified single-component, dimensionless 1D eGPE incorporating first-order LHY quantum corrections \cite{Petrov, Astrakharchik}:

\begin{eqnarray}\label{eq:QD1}
\left[i\frac{ \partial}{\partial t} =- \frac{1}{2} \frac{ \partial^2 }{\partial x^2} -  \gamma_1(t) |\psi| + \gamma_2(t) |\psi|^2 + V(x,t) + i \tau(t) \right] \psi, \nonumber\\  
\end{eqnarray}

Here, $\tau(t)$ represents the 1D time-modulated loss or gain of the condensate atoms due to the continuous escaping (dipolar relaxation) or loading (from external reservoir) of condensate atoms \cite{Yu}. The functions $\gamma_{1}(t)$ and $\gamma_{2}(t)$ are non-zero and denote the time dependent BMF and EMF coupling strengths of the binary BEC mixture. The quadratic nonlinearity reflects the 1D attractive aspect of the LHY contribution, while the cubic term addresses the conventional mean-field repulsion. In this context, $\psi$ represents the normalized macroscopic 1D condensate wavefunction. Also, the wavefunction, length, and time are measured in units of $ c_{0}=(2 \sqrt{g})^{3/2}/(\pi \xi(2|\delta g|)^{3/4}$, $\xi$, and $\hbar^{2}/m \xi^{2}$  respectively with $\xi=\pi \hbar^{2} \sqrt{|g|}/(m g \sqrt{2}g)$ is the healing length of the system \cite{Astrakharchik}. 

To find the exact analytical solution of considered 1D eGPE in equation (\ref{eq:QD1}), we start by considering an ansatz solution:
\begin{equation}\label{eq:QD2}
\psi(x,t)=\sqrt{A(t)} F[X(x,t)]Exp[i\theta(x,t) +G(t)/2],
\end{equation}
where $A(t)$ is the time-modulated amplitude and  $F[X(x,t)]$ is the space and time dependent variable akin to condensate density with $X(x,t)=A(t)[x-l(t)]$. Here, $-l(t)$ is the position of its center of mass with $l(t)=\int_{0}^{t} v(t') dt'$ \cite{Beitia}. Further,
\begin{eqnarray}\label{eq:QD3}
\theta(x,t)=a(t)-\frac{1}{2}c(t)x^{2}, 
\end{eqnarray}
represents the chosen quadratic phase form with $G(t)=\int_{0}^{t}\tau(t^{'})dt^{'}$ is the gain or loss. Here, we take $a(t)=a_{0}+\frac{\lambda_{1}-1}{2} \int_{0}^{t} A^{2}(t')dt'$ with $\lambda_{1}$ is a real constant.

Next, we substitute equation (\ref{eq:QD2}) into equation (\ref{eq:QD1}) and obtains the following real and imaginary terms:
\begin{eqnarray}\label{eq:QD4}
\frac{1}{2F}(X_{x}^{2}F_{XX}+F_{X}X_{xx})-\frac{1}{2}\theta_{x}^2-\gamma_{1}(t)e^{G(t)/2}(\sqrt{A(t)})|F| \nonumber\\
-\gamma_{2}(t)e^{ G(t)}(\sqrt{A(t)})^{2}|F|^{2}-\frac{1}{2}M(t)x^{2}-\theta_{t}=0, \\
\frac{A_{t}(t)}{2A(t)}+\frac{F_{X}X_{t}}{F}+\frac{G_{t}(t)}{2}+\frac{F_{X}X_{x}\theta_{x}}{F}+\frac{\theta_{xx}}{2}-\frac{\tau(t)}{2}=0.
\end{eqnarray}

For simplicity, we have written: $X \equiv X(x, t)$, $F \equiv F[X(x, t)]$ and $\theta \equiv \theta(x, t)$. Using the form of $\theta$, $G(t)$ and $a(t)$, we can write equation (\ref{eq:QD4}) as:
\begin{eqnarray}\label{eq:QD5}
F_{XX}- \frac{2 \gamma_{1}(t) e^{p G(t)/2}}{A^{2}(t)}(\sqrt{A(t)})|F|F -\frac{2 
\gamma_{2}(t) e^{ G(t)}}{A^{2}(t)}\nonumber\\
\times (\sqrt{A(t)})^{2}|F|^{2}F =2\lambda_{1}F,
\end{eqnarray}
with following consistency conditions:
\begin{eqnarray}\label{eq:QD6}
\frac{d l(t)}{dt}+l(t)c(t)=0,  \\
 A(t)=A_{0} exp \left[\int_{0}^{t}c(t')dt' \right], && A_{0}>0, \\
 \frac{d c(t)}{dt} -c^{2}(t)=M(t).
\end{eqnarray}

It is worth mentioning that equation (9) already in the form of the well-known Riccati equation and further with transformation $\nu(t)= e^{-\int_{0}^{t}c(t')dt'}$ allows us to write it as: 
\begin{eqnarray}\label{eq:QD7}
\nu_{tt}(t)+M(t)\nu(t)=0,
\end{eqnarray}
the well-known linear Schr\"{o}dinger equation. The correlations in equations (9) and (10) provide a significant analytical link between the droplet dynamics profile and the chosen form of the oscillator frequency $M(t)$. Since both the Schr\"{o}dinger equation and the Riccati equation can be exactly solved for various solvable quantum-mechanical systems, this connection aids in identifying the appropriate form of $M(t)$. Now, for:
\begin{eqnarray}\label{eq:QD8}
\lambda_{2}= A(t)^{(3/2)}\gamma_{1}(t) e^{G(t)/2}, && 
\lambda_{3}= A(t)\gamma_{2}(t) e^{ G(t)}, \nonumber\\
 \end{eqnarray}
we obtain equation (\ref{eq:QD5}) as:
\begin{eqnarray}\label{eq:QD9}
F_{XX}- 2 \lambda_{2}|F|F -2 \lambda_{3}|F|^{2}F =2\lambda_{1}F,
\end{eqnarray}
which is a solvable second order differential equation and represents the evolution of the droplets, for which an explicit solution can be formulated as: $F[X]=\frac{3 (\lambda_{1}/\lambda_{2}) }{1+\sqrt{1-\frac{\lambda_{1}}{\lambda_{0}} \frac{ \lambda_{3}}{ \lambda_{2}^{2}} } \cosh (\sqrt{-\lambda_{1}}X)}$ with $\lambda_{0}=-2/9$, $ \lambda_{1}<0$, $
\lambda_{2}<0$, and $\lambda_{3}>0$ \cite{Petrov,Astrakharchik}. Thus, utilizing this, we write the complete solution of the equation (\ref{eq:QD1}):

\begin{eqnarray}\label{eq:QD10}
\psi(x,t)=\sqrt{A(t)} \frac{3 (\lambda_{1}/\lambda_{2}) }{1+\sqrt{1-\frac{\lambda_{1}}{\lambda_{0}} \frac{ \lambda_{3}}{ \lambda_{2}^{2}} } \cosh (\sqrt{-\lambda_{1}}X)} \nonumber\\
\times Exp[i\theta(x,t) +G(t)/2],
\end{eqnarray}
where $X=A(t)[x-l(t)]$. By selecting $M(t)$ and $l(t)$ appropriately, we can ensure that the wave speed $v(t)$ exhibits specific desired properties. This method can be utilized to slowing or stopping, fragmenting, collapsing and reviving or even reverse the motion of droplets. In the subsequent sections, we demonstrate this for bright QDs.

\begin{figure*}[htpb]
\centering
\includegraphics[scale=0.71]{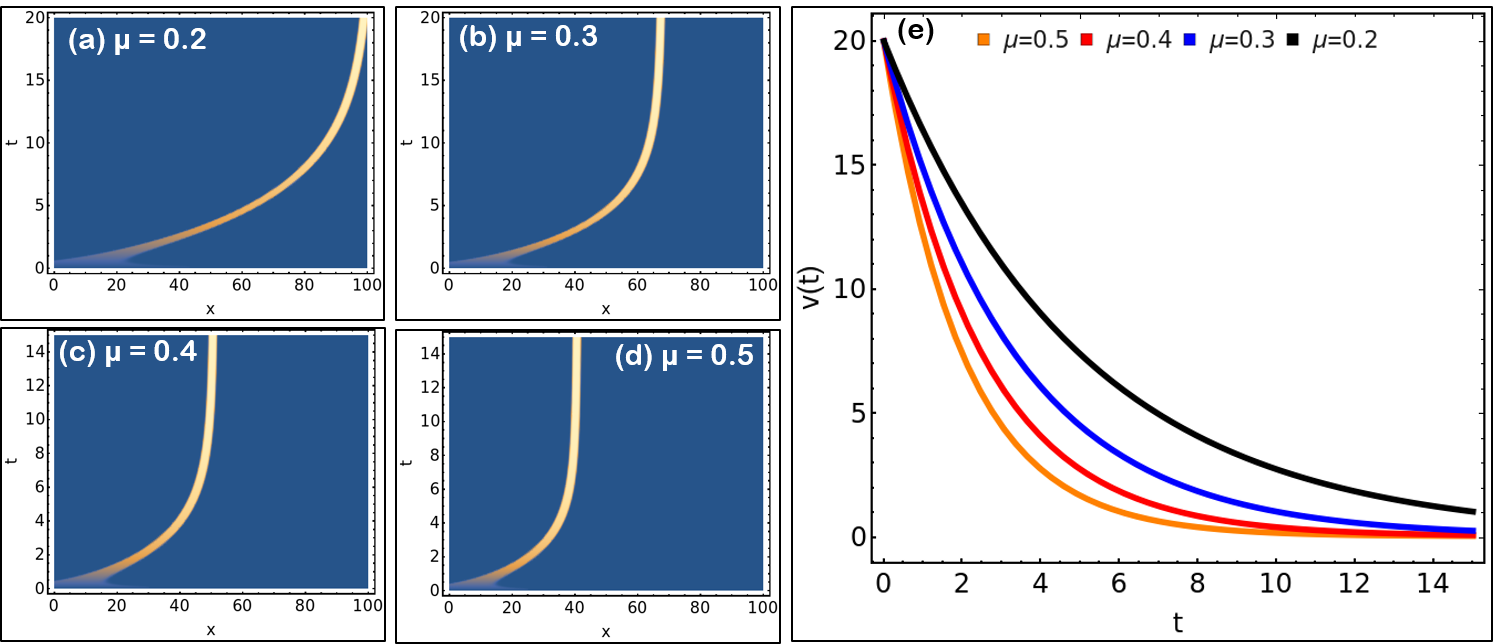}
\caption{\label{fig1} Slowing of QDs in parabolic confinement. Here, the condensate density is plotted with changing magnitude of oscillator frequency ($\mu$) for (a) $\mu=0.2$ (black), (b) $\mu=0.3$ (blue), (c) $\mu=0.4$ (red), and (d) $\mu=0.5$ (orange) along with (e) represents the corresponding velocity with respect to time is illustrated. With increase in time, the droplet speed is slowing and rate of slowing is controlled by $\mu$. For each case, $A_{0}=1$, $l_{0}=20$, $l_{\infty}=0.05$, $\lambda_{0}=-2/9$, $\lambda_{1}=-2/9$, $\lambda_{2}=-1$, $\lambda_{3}=0.999999999$ and the spatial coordinate is scaled by the oscillator length.}
\end{figure*}

\section{Slowing, stopping, or reversing, and fragmentation of Quantum Droplets}
In this section, we explore the dynamics of droplets by selecting different forms of $M(t)$. Our goal is to identify the appropriate potential forms that facilitate the speed management of droplets. As previously mentioned, the droplet peak is situated at $l(t)$, and its position change is given by $v(t)=d l(t)/dt$. To regulate this position over time, and thus control the droplet velocity, we will consider $v(t)$ as a control function. 

\subsection{Reversing of bright QDs} In order to obtain the reversing of QDs, we require that $v(t)$ changes sign at some finite time, say  $t=T>0$. For that purpose, we choose $l(t)$ as:
\begin{equation}\label{eq:QD13}
l(t)=l_0 ln\left(\frac{cosh(T)}{cosh(T-t)}\right),
\end{equation}
such that $v(t)= l_0 tanh(T- t)$, and $c(t)=\frac{tanh(T-t)}{ln\left(\frac{cosh(T)}{cosh(T-t)}\right)}$. This ensures that $v(t)>0$ for $t<T$, $v(T)=0$ and $v(t)<0$ for $t>T$. The resultant form of $M(t)$ and wavefunction can be estimated from equations (9) and (13).

Figure (\ref{fig2}) illustrates the reversing of QDs in expulsive confinement. Here, the condensate density is plotted with changing magnitude of oscillator frequency ($T$ and $l_{0}$) for (a) $T=10$, and $l_{0}=1$, (b) $T=15$, and $l_{0}=1$, (c) $T=10$, and $l_{0}=0.5$, and (d) $T=15$, and $l_{0}=0.5$. It is apparent from the figure \ref{fig2}(a)-(b) that as $T$ changes from $ 10 \rightarrow 15$, the position of droplet reversing remains same however the time changes from $10$ to $15$. In figure \ref{fig2}(a)-(c), $T$ is kept constant ($=10$) and $l_{0}$ is changed from $1 \rightarrow 0.5$ which leads to the shifting of droplet reversing position form $15$ to $7.5$ with time of reversing $10$ remaining same. Lastly, in the figure \ref{fig2}(d), both $T$ ($10 \rightarrow 15$) and $l_{0}$ ($1 \rightarrow 0.5$) are tuned leading to droplet reversal shifting from $(x=15, t=10)$ to $x=7.5, t=15$. 

\begin{figure*}[htpb]
\centering
\includegraphics[scale=0.52]{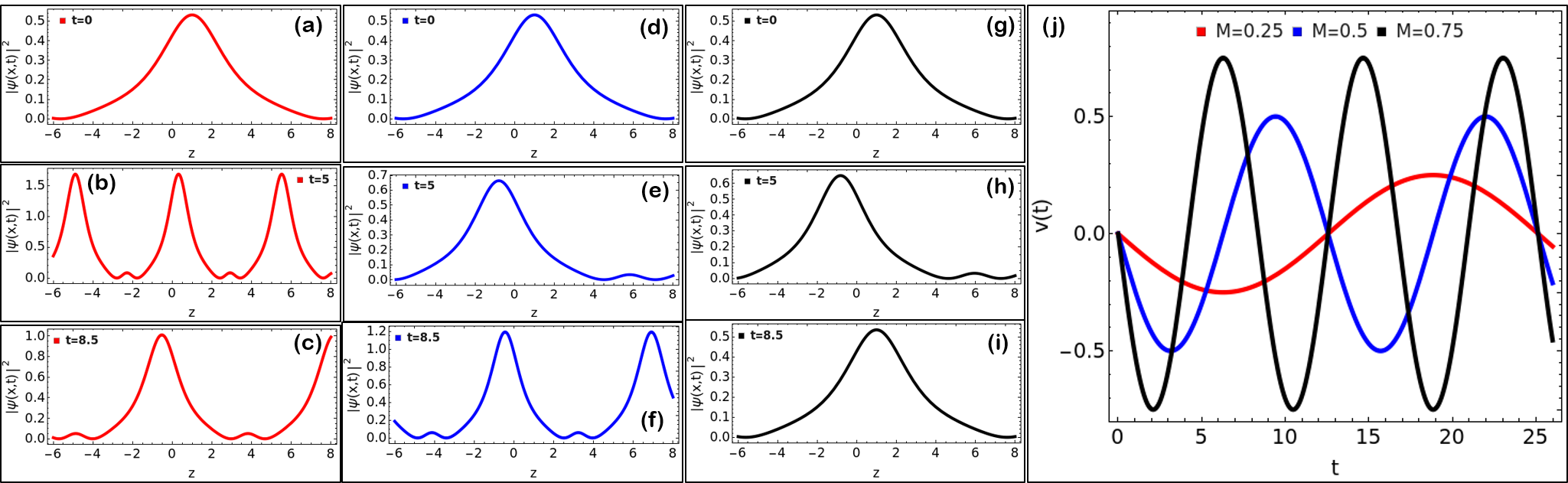}
\caption{\label{fig3} Fragmentation of QDs in parabolic confinement. Here, the condensate density is plotted with changing magnitude of oscillator frequency ($M$) for (a)-(c) $M=0.25$, (red) (d)-(f) $M=0.5$, (blue) (g)-(i) $M=0.75$ for $t=0$, $5$, and $8.5$, respectively. The corresponding velocity with respect to time for varying magnitude of (j) $M$ is illustrated. The position and time of droplet fragmentation can be tuned by oscillator frequency. For each case, $A_{0}-1.5$, $\lambda_{0}=-2/9$, $\lambda_{1}=-2/9$, $\lambda_{2}=-1$, $\lambda_{3}=0.99999999$ and the spatial coordinate is scaled by the oscillator length.}
\end{figure*}

Figure \ref{fig2}[(e)-(f)] illustrates the velocity variation with respect to time for varying magnitude of $T$ with constant $l_{0}=1$ [\ref{fig2}(e)], and varying $l_{0}$ with constant $T=15$ [\ref{fig2}(f)], respectively. It is clear from here that the position and time of droplet reversing can be tuned by oscillator frequency by modulating $T$ and $l_{0}$. For each case, $A_{0}-4$, $\lambda_{0}=-2/9$, $\lambda_{1}=-2/9$, $\lambda_{2}=-1$, $\lambda_{3}=0.99999999$.

\subsection{Slowing and stopping a bright QDs} For slowing the QDs, we require that $d l(0)/dt$ $>$ $d l(T)/dt$ with $T>0$ is some terminal time. For that purpose, we take:
\begin{equation}\label{eq:QD11}
      l(t) = l_{\infty} t + \frac{l_{0} - l_{\infty}}{\mu} (1 - e^{-\mu t}).
\end{equation}
Here, $l_{0}>l_{\infty}>0$ and $\mu>0$ are constants. Then, the velocity is given as:
\begin{equation}\label{eq:QD12}
      v(t) =  l_{\infty} + (l_0 - l_{\infty}) e^{-\mu t}, 
  \end{equation}
and from equation (7), $c(t)=v(t)/l(t)$, we obtain:
\begin{equation}\label{eq:QD13}
       c(t)= \frac{c_\infty+(c_0-c_\infty)exp(-\mu t)}{c_\infty t+\frac{c_0-c_\infty}{\mu}(1-exp(-\mu t))}.
  \end{equation}
The corresponding form of the $M(t)$ can be estimated from Riccati equation and it comes out be expulsive in nature. 

In equation \ref{eq:QD12}, $v(0)=l_{0}$, and as $t \rightarrow \infty$, then $v(t) \rightarrow l_{\infty}$ resulting in the slowing of droplets since $l_{0} > l_{\infty}$. In figure (\ref{fig1}), we illustrate the deceleration and eventual stopping of droplets for the selected form of $l(t)$. Here, the condensate density is plotted with changing magnitude of oscillator frequency ($\mu$) for (a) $\mu=0.2$ (black), (b) $\mu=0.3$ (blue), (c) $\mu=0.4$ (red), and (d) $\mu=0.5$ (orange) along with (e) represents the corresponding velocity with respect to time is illustrated. For each case, $A_{0}=1$, $l_{0}=20$, $l_{\infty}=0.05$, $\lambda_{0}=-2/9$, $\lambda_{1}=-2/9$, $\lambda_{2}=-1$, $\lambda_{3}=0.999999999$. With increase in time and the tuning of the parameter $\mu$ from $0.2 \rightarrow 0.5$ results in the slowing of the droplet speed and the rate of slowing is controlled by $\mu$. 

The parameter $\mu$ can be adjusted to control the rate of deceleration of the droplet, as $v'(0)=(l_{0}-l_{\infty})\mu$. Figure \ref{fig1}(e) shows that
$\mu$ regulates the rate of velocity change over time. 

\subsection{Fragmentation of bright QDs} The formation of bright solitary trains has been experimentally investigated by Strecker et al. in a harmonic trap \cite{Strecker}, and theoretically examined through the tuning of condensate atom gain profiles by Atre et al. \cite{Atre}. In this study, we develop an analytical framework for the fragmentation of QDs and the generation of QD trains within a regular harmonic trap. To achieve this, we choose $M(t)=M^{2}$, and express the elliptic function solution of equation (\ref{eq:QD9}) as: $F(X)= B \;\; Cn[\beta \; X,\;q]+D,$ with $D=\frac{G_{1}}{3G_{2}}$ $<0$, $\beta^{2}=-(\frac{6G_{2}}{(2q^{2}-1)})$ $D^{2}$, $B=\sqrt{\frac{2}{(2q^{2}-1)}}D$ $>0$, and $q^{2}>1/2$ \cite{Mithun1}. For this case, $A(t)=A_{0} sec(M t)$, and we can write wavefunction form utilizing equation (\ref{eq:QD10}).

Figure (\ref{fig3}) illustrates the fragmentation of QDs in parabolic confinement for $M(t) = M^{2}$ with $v(t)=-l_{0} M sin(M t)$, $l_{0}=1$. Here, the condensate density is plotted with changing magnitude of oscillator frequency ($M$) for (a)-(c) $M=0.25$, (red) (d)-(f) $M=0.5$, (blue) (g)-(i) $M=0.75$ for $t=0$, $5$, and $8.5$, respectively. For $M=0.25$, initially at $t=0$, there is single QDs. However, with increase in time from $0 \rightarrow 5 \rightarrow 8.5$, the single droplet fragments into more than three and then recombine at $t=8.5$. However, this fragmentation decreases as $M$ changes from $0.25 \rightarrow 0.5 \rightarrow 0.75$. The corresponding velocity with respect to time for varying magnitude of (j) $M$ is illustrated. It is apparent from figure \ref{fig3}(j), with increasing $M$, the velocity of droplet is increasing along with its period of oscillation with time leading to increase in the acceleration.  The position and time of droplet fragmentation can be tuned by oscillator frequency. For each case, $A_{0}-1.5$, $\lambda_{2}=-1$, $\lambda_{3}=0.99999999$.

\begin{figure*}[htpb]
\centering
\includegraphics[scale=1.02]{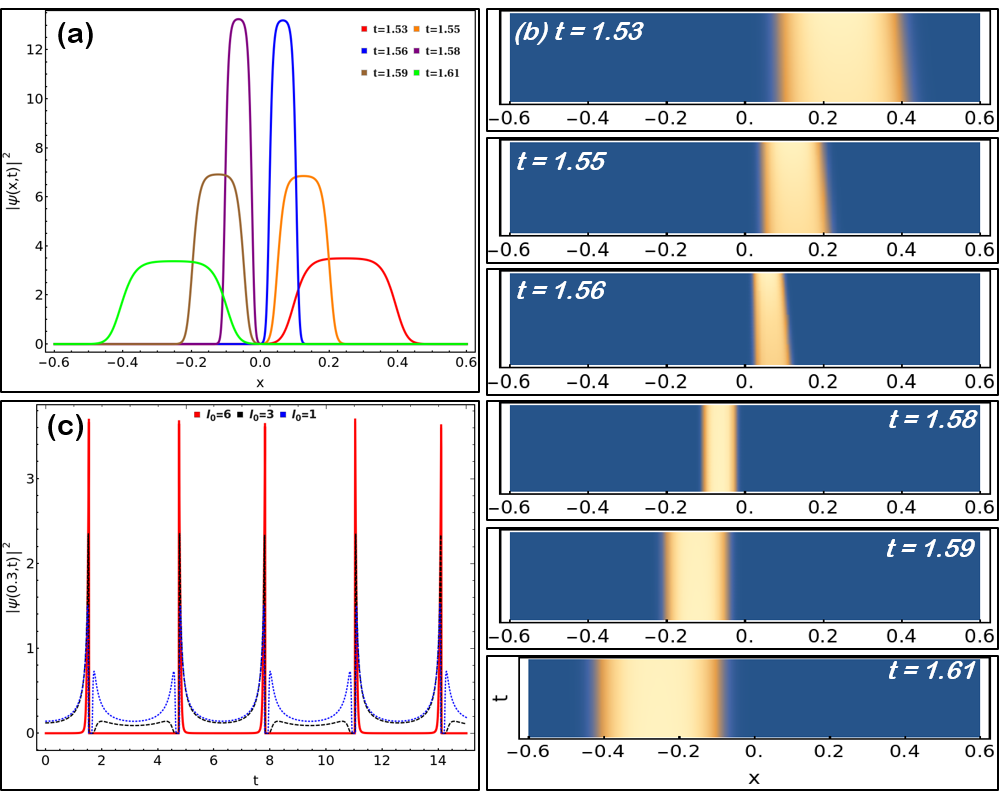}
\caption{\label{fig4} Collapse and revival of QDs: Here, the condensate density is plotted with changing magnitude of time in (a)-(b) at $l(t)=l_{0} cos(M_{0}t$ to illustrated the collapse and revival of droplets for $1.53$ (red), $1.55$ (orange), $1.56$ (blue), $1.58$ (purple), $1.59$ (brown), and $t=1.61$ (green), respectively with $l_{0}=6$.  (c) Variation of condensate density with changing magnitude of droplet velocity $v$ at $x=0.3$ is depicted for $l_{0}=6$ (red), $3$ (black dashed) and $1$ (blue dotted), respectively. For each case, $A_{0}=4$, $M_{0}=1$, $\lambda_{1}=-2/9$, $\lambda_{2}=-1$, $\lambda_{3}=0.9999$ and the spatial coordinate is scaled by the oscillator length.}
\end{figure*}

\subsection{Collapse and revival of bright QDs} In this section, we investigate the collapse and revival of QDs in the presence of regular parabolic trap by choosing $M(t)= M_{0}^2$ and $l(t)=l_{0} cos(M_{0}t)$. For the choosen parameters, the wavefunction takes the form: $\psi(x,t)=\sqrt{A_{0}(t)sec(M_{0}t)} \frac{3 (\lambda_{1}/\lambda_{2}) }{1+\sqrt{1-\frac{\lambda_{1}}{\lambda_{0}} \frac{ \lambda_{3}}{ \lambda_{2}^{2}} } \cosh (\sqrt{-\lambda_{1}}(x-l{0} cos(M_{0}t)))} \times Exp[i\theta(x,t) +G(t)/2]$. Figure (\ref{fig4}) depicts the collapse and revival of droplets for given trap configuration. Here, the condensate density is plotted with changing magnitude of time in (a)-(b) with $l(t)=l_{0} cos(M_{0}t$ to illustrated the collapse and revival of droplets for $1.53$ (red), $1.55$ (orange), $1.56$ (blue), $1.58$ (purple), $1.59$ (brown), and $t=1.61$ (green), respectively with $l_{0}=6$.  For each case, $A_{0}=4$, $M_{0}=1$, $\lambda_{1}=-2/9$, $\lambda_{2}=-1$, $\lambda_{3}=0.9999$ and the spatial coordinate is scaled by the oscillator length.

It is apparent from figure \ref{fig4}(a)-(b) that as $t$ tends from $1.53 \rightarrow 1.61$, the droplet location is at $x=0.3$ and it starts reviving. The droplet regains the same maxima magnitude at $t=1.61$ near $x=-0.3$. It is evident from the figure that the center of mass of droplet is oscillating and illustrating the collapse and revival. In figure, the peak value of condensate density is observed at $t=1.56, 1.58$ and its half-maximum is exists at $t=1.55, 1.59$. Further, figure \ref{fig4})(c) represents the variation of condensate density with changing magnitude of droplet velocity $v$ at $x=0.3$ for $l_{0}=6$ (red), $3$ (black dashed) and $1$ (blue dotted), respectively. The tuning of $l_{0}$ provides control on the rate of collapse with respect to time at the given position. Additionally, the collapse and revival time is a function of $(n+1/2) \pi$ with $n=0,1,2,3..$. Here, the harmonic trap is coupled with the center of mass due to Kohn’s theorem.  

\begin{figure*}[htpb]
\centering
\includegraphics[scale=1.36]{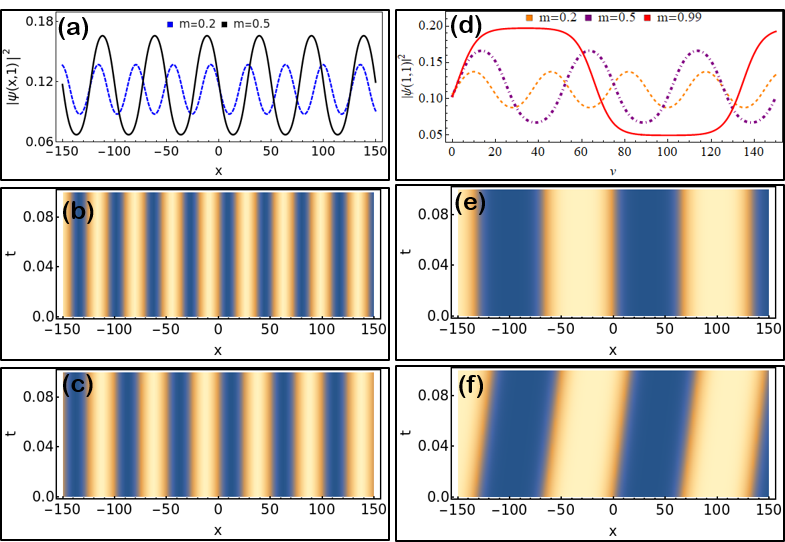}
\caption{\label{fig5} Supersolidity of QDs: Here, the condensate density is plotted with changing magnitude of modulus parameter ($m$) for (a) $m=0.2$ and $m=0.5$ at $t=1$ and corresponding density plots are illustrated for (b) $m=0.2$, and (c) $m=0.5$, respectively.  (d) Variation of condensate density with changing magnitude of droplet velocity $v$ at $x=1, t=1$ is depicted for $m=0.2$, $0.5$ and $0.9$. The density variation with tuning of $v$ is represented for (e) $m=0.99$, $v=0$, and  (f) $m=0.99$, $v=150$, respectively.  For each case, $\lambda_{1}=-2/9$, $\lambda_{2}=-1$, $\lambda_{3}=0.9999$ and the spatial coordinate is scaled by the oscillator length.}
\end{figure*}

\begin{figure*}[htpb]
\centering
\includegraphics[scale=0.79]{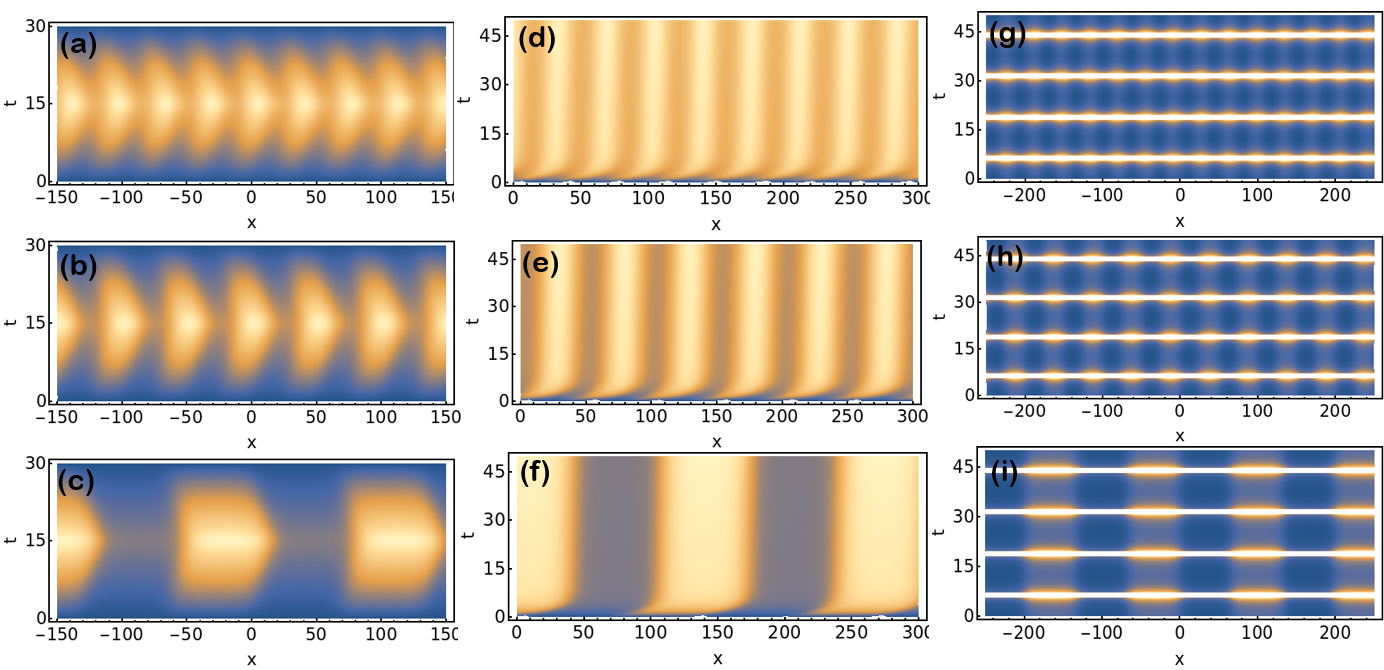}
\caption{\label{fig6} Supersolidity of QDs: Figure illustrate the supersolid formation for cases (a)-(c) reversing for $T=15$, $l_{0}=1$, $A_{0}=4$, and (d)-(f) slowing and stopping for $\mu=0.5$, $A_{0}=1$, $l_{0}=20$, $l_{\infty}=0.05$, and (g)-(i) collapse and revival for $A_{0}=1.5$, $M_{0}=0.25$, $l_{0}=0.25$ of QDs. Here, the condensate density is plotted with changing magnitude of modulus parameter ($m$) for (a),(d), (g) $m=0.2$, (b),(e), (h) $m=0.5$, and (c),(f), (i) $m=0.99$, respectively. In this, $\lambda_{1}=-2/9$, $\lambda_{2}=-1$, $\lambda_{3}=0.9999$ and the spatial coordinate is scaled by the oscillator length.}
\end{figure*}

\section{Supersolid phase in chosen parabolic trap}
A supersolid is an exotic state of matter characterized by a crystalline structure coexisting with a superfluid component, which exhibits zero viscosity and flows without resistance. In this section, we investigate the supersolid behavior in QDs for the above-discussed cases of reversing, slowing or stopping, and collapsing and reviving. For that purpose, we consider the equation (\ref{eq:QD9}): $F_{XX}- 2 \lambda_{2}|F|F -2 \lambda_{3}|F|^{2}F =2\lambda_{1}F,$ with $\lambda_{2}<0$, $\lambda_{3}>0$, and take a propagating periodic ansatz solution \cite{Parit}:

\begin{eqnarray}\label{eq:QD13}
F[(x-v t/\xi)]=A+ B \times sn \left[ \frac{x-vt}{\xi}, m\right],
\end{eqnarray}

with $m$ as the modulus parameter of the elliptic function, we calculate the healing length as $\xi^{2} = \frac{(m+1)}{2} \frac{9 \lambda_{3}}{\lambda_{2}^{2}}$. For a moving supersolid, the chemical potential is bounded below with $\lambda_{1}(min)=-\frac{2 \lambda_{2}^{2}}{9 \lambda_{3}} <0$, which is identical to that of the self-trapped droplet \cite{Petrov}. The chosen periodic solution is accompanied by a positive background $A=\frac{\lambda_{2}}{3 \lambda_{3}}$, which is one-third the value of the uniform background. It is important to note that the supersolid solutions do not smoothly transition to a constant condensate in the limit $m \rightarrow 0$ when $B=0$. The amplitude $B=\pm \sqrt{\frac{2m}{(m+1)}} A$, never exceeds that of the background within the allowed energy range of the modulus parameter $0<m<1$. Here, $F_{min}/F_{max}=A \pm B=A(1 \pm \sqrt{\frac{2n}{m+1}})>0$, clearly showing that for the quantum supersolid immersed in a residual BEC.

Figure (\ref{fig5}) illustrates the supersolid phase under the free space scenario using equation (\ref{eq:QD9}). Here, the condensate density is plotted with changing magnitude of modulus parameter ($m$) for (a) $m=0.2$ and $m=0.5$ at $t=1$ and corresponding density plots are illustrated for (b) $m=0.2$, and (c) $m=0.5$, respectively. For each case, $\lambda_{1}=-2/9$, $\lambda_{2}=-1$, $\lambda_{3}=0.9999$ and the spatial coordinate is scaled by the oscillator length. It is apparent from the figure that the number of droplets is correlated with the inter-droplet spacing (i.e. peak to peak distance). With $m$ tending from $0.2 \rightarrow 0.5$ leads to decrease in inter-droplet spacing and number of droplets, however, the intensity increase with increasing $m$. Further, we investigate the droplet dynamics at $x=1, t=1$ with changing velocity in figure \ref{fig5}(d) for $m=0.2,$ $0.5$, and $0.99$. It is evident from the figure that the droplet size and inter-droplet distance is increasing with increase in the velocity. We plotted the density variation with tuning of $v$ is represented for (e) $m=0.99$, $v=0$, and  (f) $m=0.99$, $v=150$, respectively.

\subsection{Supersolid phase for reversing droplet} In this case, we investigate the supersolid phase for the slowing of QDs in present of parabolic trap. For that we take $l(t)=l_0 ln\left(\frac{cosh(T)}{cosh(T-t)}\right)$, and $M(t) \neq 0$. Then, from equations (\ref{eq:QD10}) and (\ref{eq:QD13}), the wavefunction becomes:
\begin{eqnarray}\label{eq:QD14}
\hspace{-0.5cm} \psi(x,t)= \left[ \frac{\lambda_{2}}{3 \lambda_{3}} +\frac{2}{9}\frac{\lambda_{2}^{2}}{3 \lambda_{3}^{2}}\frac{m}{m+1} sn \left(\frac{x-l_0 ln \frac{cosh(T)}{cosh(T-t)}}{\xi}, m \right) \right] \nonumber\\
\times \sqrt{A_{0}Log[K sech(t-T)]} Exp[i\theta(x,t) +G(t)/2].\nonumber\\
\end{eqnarray}

Utilizing the wavefunction form in equation (\ref{eq:QD14}), in figure \ref{fig6} [(a)-(c)] illustrate the supersolid formation for reversing QDs for $T=15$, $l_{0}=1$, $A_{0}=4$, $\lambda_{1}=-2/9$, $\lambda_{2}=-1$, and $\lambda_{3}=0.9999$. Here, the condensate density is plotted with changing magnitude of modulus parameter $m$ for (a) $m=0.2$, (b) $m=0.5$, and (c) $m=0.99$, respectively. As $T=15$, so reversing is evident at $t=15$ and with increase in the magnitude of $m$ from $0.2 \rightarrow 0.99$ leads increase in inter-droplet spacing i.e. lattice period and decrease in number of droplets i.e lattice sites of the formed supersolid. 

\subsection{Supersolid phase for slowing and stopping of droplet} Here, we emphasize on the supersolid phase formation for the case of slowing and stopping of QDs by choosing $l(t) = l_{\infty} t + \frac{l_{0} - l_{\infty}}{\mu} (1 - e^{-\mu t})$ and $M(t) \neq 0$ i.e. parabolic trap. For this scenario, the wavefunction constructed  from equations (\ref{eq:QD10}) and (\ref{eq:QD13}) is $ \psi(x,t)= \left[ \frac{\lambda_{2}}{3 \lambda_{3}} +\frac{2}{9}\frac{\lambda_{2}^{2}}{3 \lambda_{3}^{2}}\frac{m}{m+1} sn \left(\frac{x-l_{\infty} t - \frac{l_{0} - l_{\infty}}{\mu} (1 - e^{-\mu t})}{\xi}, m \right) \right] 
\times \sqrt{A_{0} [l_{0}- l_{0} Exp[-\mu t] + l_{\infty}(-1+Exp[-\mu t]+ \mu t)]} \times Exp[i\theta(x,t) +G(t)/2].$

\begin{figure}[htpb]
\centering
\includegraphics[width=\linewidth]{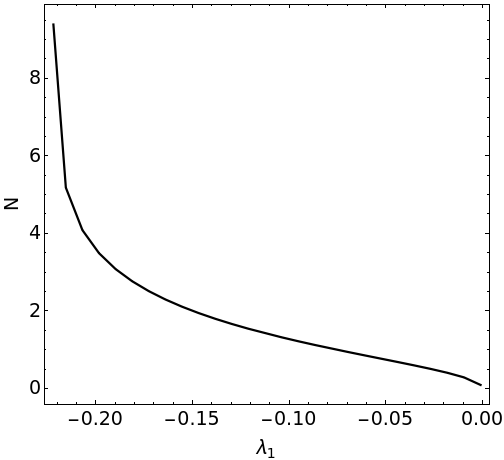}
\caption{\label{fig7} Stability plots for $N$ versus $\lambda_{1}$}. 
\end{figure}

Figure \ref{fig6} [(d)-(f)] depicts the formation of supersolid phase for the slowing and stopping of QDs in parabolic trap with varying magnitude of elliptic modulus parameter $m$. Here, $\mu=0.5$, $A_{0}=1$, $l_{0}=20$, $l_{\infty}=0.05$, $\lambda_{1}=-2/9$, $\lambda_{2}=-1$, $\lambda_{3}=0.9999$. It is apparent from the figure that as $m$ tends from $0.2 \rightarrow 0.99$, the number of droplets decreases and the inter-droplet peak to peak distance increases. Further, initially the droplet position is changing with time but gradually it becomes stationary representing stopping of droplet. 

\subsection{Supersolid phase for collapse and revival of droplet} Next, we consider the formation of supersolid phase in the case of collapse and revival of QDs. For that, we take $l(t)=l_{0} cos(M_{0} t)$, and $M(t)=M_{0}^2$ i.e. $V(x,t) =(1/2)M_{0}^2 x^2 $. This results in the form of wavefunction: $\psi(x,t)= \left[ \frac{\lambda_{2}}{3 \lambda_{3}} +\frac{2}{9}\frac{\lambda_{2}^{2}}{3 \lambda_{3}^{2}}\frac{m}{m+1} sn \left(\frac{x-l_{0} cos(M_{0} t)}{\xi}, m \right) \right] \nonumber\\
\times \sqrt{A_{0} sec(M_{0}t)} Exp[i\theta(x,t) +G(t)/2].$

Like previous cases, we plot the condensate density with changing magnitude of elliptic parameter $m$ for the given case in figure \ref{fig6} [(g)-(i)]. Here, for $m=0.2$, the number of droplets are more and their inter spacing is less in comparison to $m=0.99$. Further, the collapse and revival of these droplets is evident from figure with increasing time.

\section{Stability Analysis} In the previous sections, we demonstrated the droplet speed management and supersolid behavior under the external parabolic confinement utilizing the localized and periodic solutions of equation (\ref{eq:QD9}). The localized droplet solution of equation (\ref{eq:QD9}): $F[X]=\frac{3 (\lambda_{1}/\lambda_{2}) }{1+\sqrt{1-\frac{\lambda_{1}}{\lambda_{0}} \frac{ \lambda_{3}}{ \lambda_{2}^{2}} } \cosh (\sqrt{-\lambda_{1}}X)}$, with $\lambda_{0}=-2/9$, exists in the interval of $\lambda_{0}<\lambda_{1}<-2/9$ for $ \lambda_{1}<0$, $\lambda_{2}<0$, and $\lambda_{3}>0$ \cite{Petrov,Astrakharchik}. The scaled number of atoms in the droplet for wavefunction of the form given in equation (\ref{eq:QD10}), $N=\int_{-\infty}^{+\infty}|\psi|^2 \partial x$, one can estimate the correlation between normalization $N$ and $\lambda_{1}$ as: 
 \begin{eqnarray}\label{eq:QD15}
N=\frac{4}{3} \left[ ln \left( \frac{1+ \sqrt{\frac{\lambda_{1}}{\lambda_{0}}}}{1- \sqrt{\frac{\lambda_{1}}{\lambda_{0}}}}\right) -\sqrt{\frac{\lambda_{1}}{\lambda_{0}}}\right].
\end{eqnarray}

Equation (\ref{eq:QD15}) calculates the magnitude of $N$ in the presence of parabolic trap and matches the value of $N$ reported for free space \cite{Astrakharchik,Tylutki}. Therefore, even with an external trap, the system maintains continuous symmetry, and $N$ is conserved according to Noether's theorem \cite{Sardanashvily}.

Next, we conduct a linear stability analysis of the second periodic solution of equation (\ref{eq:QD9}) by considering the wavefunction form of equation (\ref{eq:QD14}).  In the context of linear stability, our initial goal is to evaluate the modulational instability (MI), which uncovers the regions of instability within the parameter space using Bogoliubov-de Gennes (BdG) linearized equations \cite{Mithun1}. For that purpose, we consider the applied perturbation as plane waves and investigate the BdG spectrum. 

Here, we introduce a small perturbation to the stationary solution, expressed as $\psi(x,t) = \psi_{0}(x) +\delta \psi(x,t)$, with $\delta \psi(x,t) << 1$. Now, substituting $\psi(x,t)$ in equation (\ref{eq:QD1}), and linearizing it, we write the eigenvalue equation in terms of the perturbation $\delta \psi$:

\begin{eqnarray}\label{eq:QD16}
i\frac{ \partial \delta \psi}{\partial t} &=& - \frac{1}{2} \frac{ \partial^2 \delta \psi}{\partial x^2} -  \gamma_1 (2 \sqrt{n} \delta \psi) \nonumber\\ 
&&+ \gamma_2 (2n \delta \psi + n \delta \psi^{*}) + V(x) \delta \psi
\end{eqnarray}
with $n$ =$|\psi_{0}(x)|^{2}$, and $ \tau(t) =0$. Further breaking down $\delta \psi(x,t)=  \begin{bmatrix} 
	\delta \psi_{R}  \\
	\delta \psi_{I}
\end{bmatrix} $ into its real and imaginary parts transforms equation (\ref{eq:QD16}) into the familiar BdG equation:

 \begin{eqnarray}\label{eq:QD17}
 \begin{bmatrix} 
	 -\frac{1}{2}\frac{\partial^{2}}{\partial x^{2}} + p_{1} & 0  \\
	0 & -\frac{1}{2}\frac{\partial^{2}}{\partial x^{2}} +  p_{2} \\
	\end{bmatrix} \begin{bmatrix} 
	 \delta \psi_{R}  \\
	\delta \psi_{I} \\
	\end{bmatrix}=
 \begin{bmatrix} 
	 0 & 1  \\
	-1 & 0 \\
	\end{bmatrix} \frac{\partial}{\partial t}\begin{bmatrix} 
	 \delta \psi_{R}  \\
	\delta \psi_{I} \\
	\end{bmatrix}, \nonumber\\
 \end{eqnarray}
where $\delta \psi_{R}$, $\delta \psi_{I}$ are real and imaginary components of the small perturbation $\delta \psi(x,t)$. Also, $p_{1}=3 n \gamma_{2} -2 \gamma_{1} n^{1/2} + V(x)$, and $p_{2}=\gamma_{2} n - 2 \gamma_{1} n^{1/2} + V(x)$. Assuming $\delta \psi = exp[i (q x-\omega t)]$ and substituting it into equation (\ref{eq:QD17}), one can derive the eigenmodes of the perturbation, where $q$ represents the wave number and $\omega$ denotes the frequency. The resulting dispersion relation can be characterized as: $\omega^2=\frac{q^4}{4} + q^2 \left(2 n \gamma_{2} - 2 \gamma_{1} n^{1/2} \right),$ by neglecting the $q$ independent terms in the dispersion relation. Therefore, one can conclude that the instability will occur when $2 n \gamma_{2}<2 \gamma_{1} n^{1/2}$ and $q^2 \left(2 \gamma_{1} n^{1/2} -2 n \gamma_{2} \right)> \frac{q^4}{4}$.

\section{Conclusion}
In summary, this work presents a novel management method for controlling the speed and direction of self-bound quantum droplets (QDs) within a binary Bose-Einstein condensate mixture subjected to time-modulated external harmonic confinement. By employing the 1D eGPE, QDs were constructed under both regular and expulsive parabolic traps with temporally varying attractive quadratic beyond mean-field and repulsive cubic mean-field atom-atom interactions. The study demonstrates the ability to slow down, stop, reverse, collapse and revival, and induce fragmentation's in the droplets. Additionally, the solutions reveal crystalline order with a superfluid background, indicative of supersolid behavior across different parameter domains. Notably, one-third of the constant background corresponds precisely to the lowest residual condensate. Here, the QDs form a periodic array with a constant inter-droplet spacing, immersed in a residual BEC background, which establishes inter-droplet coherence. These findings have potential applications in matter-wave interferometry and quantum information processing, providing a new avenue for precise control in quantum systems.

\begin{center}
    {\bf Acknowledgement}
\end{center}
AN acknowledge insightful discussion with Prof. Prasanta K Panigraphi in preparation of the manuscript.


\begin{thebibliography}{99}
\bibitem{Luo} Z.-H. Luo,  W. Pang, B. Liu, Y.-Y. Li and B. A. Malomed. A new form of liquid matter: Quantum droplets. {\it Frontiers of Physics} \textbf{16}, 32201 (2021).

\bibitem{ttcher} F. B\"{o}ttcher, J.-N. Schmidt, J. Hertkorn, K. Ng, S. Graham, M. Guo, T. Langen, and T. Pfau,  New states of matter with fine-tuned interactions: quantum droplets and dipolar supersolids. {\it Reports on Progress in Physics} {\bf 84}, 012403 (2021).

\bibitem{Malomed1} B. A. Malomed, The family of quantum droplets keeps expanding, Front. Phys. {\bf 16}, 22504 (2021).

\bibitem{Khan1} A. Khan and A. Debnath, Frontiers in Physics {\bf 10}, 534 (2022).

\bibitem{Rakshit} D. Rakshit, T. Karpiuk, M. Brewczyk, and M. Gajda, Quantum Bose-Fermi droplets, {\it SciPost Phys.} {\bf 6}, 079 (2019).

\bibitem{Lee} T. D. Lee, K. Huang, and C. N. Yang, Physical Review
{\bf 106}, 1135 (1957).

\bibitem{Petrov} D. S. Petrov, Quantum Mechanical Stabilization of a Collapsing Bose-Bose Mixture, Phys. Rev. Lett. {\bf 115}, 155302 (2015).

\bibitem{Petrov1} D. S. Petrov and G. E. Astrakharchik, Ultradilute lowdimensional liquids, Phys. Rev. Lett. {\bf 117}, 100401 (2016).

\bibitem{Kadau} H. Kadau, M. Schmitt, M. Wenzel, C. Wink, T. Maier, I. FerrierBarbut, and T. Pfau, Observing the Rosensweig instability of a quantum ferrofluid, Nature (London) {\bf 530}, 194 (2016).

\bibitem{Schmitt} M. Schmitt, M. Wenzel, F. Böttcher, I. Ferrier-Barbut, and T. Pfau, Self-bound droplets of a dilute magnetic quantum liquid, Nature (London) {\bf 539}, 259 (2016).

\bibitem{Schmitt1} F. B\"{o}ttcher, J.-N. Schmidt, M. Wenzel, J. Hertkorn, M. Guo, T. Langen, and T. Pfau, Transient Supersolid Properties in an Array of Dipolar Quantum Droplets, Phys. Rev. X {\bf 9}, 011051 (2019).

\bibitem{Cabrera} C. R. Cabrera, L. Tanzi, J. Sanz, B. Naylor, P. Thomas, P. Cheiney, and L. Tarruell, Quantum liquid droplets in a mixture of Bose-Einstein condensates, Science {\bf 359}, 301 (2018).

\bibitem{Cheiney} P. Cheiney, C. R. Cabrera, J. Sanz, B. Naylor, L. Tanzi, and L. Tarruell, Bright Soliton to Quantum Droplet Transition in a Mixture of Bose-Einstein Condensates, Phys. Rev. Lett. {\bf 120}, 135301 (2018).

\bibitem{Semeghini} G. Semeghini, G. Ferioli, L. Masi, C. Mazzinghi, L. Wolswijk, F. Minardi, M. Modugno, G. Modugno, M. Inguscio, and M. Fattori, Self-Bound Quantum Droplets of Atomic Mixtures in Free Space, Phys. Rev. Lett. 120, 235301 (2018).

\bibitem{Astrakharchik} G. E. Astrakharchik and B. A. Malomed, Dynamics of one-dimensional quantum droplets, Phys. Rev. A {\bf 98}, 013631 (2018).

\bibitem{Ivan} I. Morera, G. E. Astrakharchik, A. Polls, and B. Julia-Diaz, Universal Dimerized Quantum Droplets in a One-Dimensional Lattice, {\it Phys. Rev. Lett.} {\bf 126}, 023001 (2021).

\bibitem{Ivan1} I. Morera, G. E. Astrakharchik, A. Polls, and B. Julia-Diaz, Quantum droplets of bosonic mixtures in a one-dimensional optical lattice, Phys. Rev. Research {\bf 2}, 022008(R) (2020).

\bibitem{Maitri2} M. R. Pathak, and A. Nath. Droplet to soliton crossover 
at negative temperature in presence of bi‑periodic optical lattices.  Scientific Reports {\bf 12}, 6904 (2022).

\bibitem{Zin} P. Zin, M. Pylak, T. Wasak, M. Gajda, and Z. Idziaszek, Quantum Bose-Bose droplets at a dimensional crossover, Phys. Rev. A {\bf 98}, 051603(R) (2018).

\bibitem{Lavoine} L. Lavoine and T. Bourdel, Beyond-mean-field crossover from one dimension to three dimensions in quantum droplets of binary mixtures, Phys. Rev. A {\bf 103}, 033312, (2021).

\bibitem{Tylutki} M. Tylutki, G. E. Astrakharchik, B. A. Malomed, and D. S. Petrov, Collective excitations of a one-dimensional quantum droplet, Phys. Rev. A  {\bf 101}, 051601(R) (2020).


\bibitem{Hu} H. Hu and X.-J. Liu, Collective excitations of a spherical ultradilute quantum droplet, Phys. Rev. A {\bf 102}, 053303, (2020).

\bibitem{Edmonds} M. Edmonds, Dark quantum droplets and solitary waves in beyond-mean-field Bose-Einstein condensate mixtures, Phys. Rev. Research {\bf 5}, 023175 (2023).

\bibitem{Zhang} X. L. Zhang, X. X. Xu, Y. Y. Zheng, Z. P. Chen, B. Liu, C. Q. Huang, B. A. Malomed, Y. Y. Li, Semidiscrete quantum droplets and vortices, Phys. Rev. Lett. {\bf 123}, 133901 (2019).


\bibitem{Xia} X. Hu, Z. Li, Y. Guo, Y. Chen, and X. Luo, Scattering of one-dimensional quantum droplets by a reflectionless potential well, Phys. Rev. A {\bf 108}, 053306, (2023).

\bibitem{Elhad} K. M. Elhadj, L. Al Sakkaf, A. Boudjemâa, U. Al Khawaja, Quantum droplet molecules in Bose-Bose mixtures, Physics Letters A, {\bf 494}, 129274, (2024).

\bibitem{Parit} M. K. Parit, G. Tyagi, D. Singh and P. K Panigrahi, Supersolid behavior in one-dimensional self-trapped Bose–Einstein condensate, J. Phys. B: At. Mol. Opt. Phys. {\bf 54} 105001, (2021).

\bibitem{Usama} U. A. Khawaja and L. A. Sakkaf {\it Handbook of Exact Solutions to the Nonlinear Schrödinger Equations} (IOP Publishing Ltd) 2019.

\bibitem{Dalfovo} F. Dalfovo, S. Giorgini, L. P. Pitaevskii, and S. Stringari, Theory of Bose-Einstein condensation in trapped gases, Rev. Mod. Phys. {\bf 71} 463 (1999).

\bibitem{Pethick} C. J. Pethick  and H. Smith {\it Bose-Einstein Condensation in Dilute Gases} (Cambridge: Cambridge University Press) 2008.

\bibitem{Atre}  R. Atre,  P. K. Panigrahi and G. S. Agarwal, Class of solitary wave solutions of the one-dimensional Gross-Pitaevskii equation, Phys. Rev. E \textbf{73}, 056611 (2006).

\bibitem{Beitia} J. Belmonte-Beitia, V. M. Pérez-Garcia, V. Vekslerchik, and V. V. Konotop, Localized Nonlinear Waves in Systems with Time- and Space-Modulated Nonlinearities, Phys. Rev. Lett. {\bf 100} 164102 (2008).

\bibitem{Sulem} C. Sulem and P. L. Sulem {\it The Nonlinear Schr\"{o}dinger Equation: Self-Focussing and Wave Collapse} (New York: Springer) 1999.

\bibitem{Rajaraman} R. Rajaraman  {\it Solitons and Instantons} (Amsterdam: North-Holland) 1982. 

\bibitem{Beitia1} J. Belmonte-Beitia, J. Cuevas, Solitons for the cubic–quintic nonlinear Schrödinger equation with time- and space-modulated coefficients, J. Phys. A: Math. Gen. {\bf 42} 165201, 2009.

\bibitem{Wang} C.-Y. Wang, The analytic solutions of Schrödinger equation with Cubic-Quintic nonlinearities, Results in Physics {\bf 10} 150–154, 2018.

\bibitem{Ajay} A. Nath, and U. Roy, A unified model for an external trap in a cigar-shaped Bose-Einstein condensate, J.Phys. A: Mathematical and Theoretical, \textbf{47}, 415301, 2014.

\bibitem{Roy} U. Roy, B. Shah, K. Abhinav and P. K. Panigrahi, Gapped solitons and periodic excitations in strongly coupled BECs, J. Phys. B: At. Mol. Opt. Phys. {\bf 44}, 035302 (2011).

\bibitem{Tang} X. Y. Tang  and P. K. Shukla, Solution of the one-dimensional spatially inhomogeneous cubic-quintic nonlinear Schrödinger equation with an external potential, Phys. Rev. A {\bf 76} 013612 (2007).

\bibitem{Baines} L. W. S. Baines and R. A. Van Gorder, Soliton wave-speed management: Slowing, stopping, or reversing a solitary wave, Phys. Rev. A 97, 063814, (2018).

\bibitem{Serkin} V. N. Serkin  and A. Hasegawa, Novel Soliton Solutions of the Nonlinear Schrödinger Equation Model, Phys. Rev. Lett. {\bf 85}, 4502 (2000).

\bibitem{Kruglov} V. I. Kruglov, M. K. Olsen  and M. J. Collett, Quantum and thermal fluctuations of trapped Bose-Einstein condensates, Phys. Rev. A {\bf 72}, 033604 (2005).

\bibitem{Bisset} R. N. Bisset, R. M. Wilson, D. Baillie, and P. B. Blakie, Ground-state phase diagram of a dipolar condensate with quantum fluctuations, 
Phys. Rev. A {\bf 94}, 033619 (2016).

\bibitem{Cinti}  F. Cinti, A. Cappellaro, L. Salasnich, and T. Macri, Superfluid Filaments of Dipolar Bosons in Free Space, Phys. Rev. Lett. {\bf 119}, 215302 (2017).

\bibitem{Macia} A. Macia, J. S´anchez-Baena, J. Boronat, and F. Mazzanti, Droplets of Trapped Quantum Dipolar Bosons, Phys. Rev. Lett. {\bf 117}, 205301 (2016).

\bibitem{Boronat} V. Cikojević, K. Dželalija, P. Stipanović, L. Vranješ Markić, and J. Boronat, Ultradilute quantum liquid drops, Phys. Rev. B {\bf 97}, 140502(R), (2018).

\bibitem{Maitri} M. R. Pathak, and A. Nath. Dynamics of Quantum Droplets in an External Harmonic Confinement.  {\it Scientific Reports} {\bf 12}, 6904 (2022).

\bibitem{Bhatia} S Bhatia, CN Kumar, A Nath, Investigation of one-dimensional quantum droplets in a temporally perturbed external harmonic trap, Physics Letters A {\bf 492}, 129228 (2023).

\bibitem{Kartashov} Y. V. Kartashov, D. A. Zezyulin, Enhanced mobility of quantum droplets in periodic lattices, Chaos, Solitons and Fractals, {\bf 182}, 114838, (2024).

\bibitem{Maitri1} M. R. Pathak, and A. Nath. Formation of Matter-Wave Droplet Lattices in Multi-Color Periodic Confinements. {\it Symmetry} {\bf 14}, 963 (2022).

\bibitem{Shukla} A. Shukla, Neeraj and P. K Panigrahi, Kink-like solitons in quantum droplet, J. Phys. B: At. Mol. Opt. Phys. {\bf 54} 165301, (2021).

\bibitem{Mithun1}T. Mithun, A. Maluckov, K. Kasamatsu, B. A. Malomed, and A. Khare, Symmetry {\bf 12}, 174 (2020).

\bibitem{Hau} Hau, L.Y., Harris, S.E., Dutton, Z., and Behroozi, C.H., Light speed reduction to 17 metres persecond in an ultracold atomic gas. Nature
{\bf 397}, 594-598 (1999)

\bibitem{Dutton} Dutton, Z., Budde, M., Slowe, C., and Hau, L.v., Observation of
quantum shock waves created with ultra-compressed slow light pulses in a Bose-Einstein Condensate. Science {\bf 293}, 663-668 (2001).

\bibitem{Kengne} Kengne, E.; Liu, W.-M.; Malomed, B.A. Spatiotemporal engineering of matter-wave solitons in Bose–Einstein condensates. Phys. Rep. {\bf 899}, 1-62, (2021).

\bibitem{Yu} Yu. Kagan, A. E. Muryshev, and G. V. Shlyapnikov, Collapse and Bose-Einstein Condensation in a Trapped Bose Gas with Negative Scattering Length, Phys. Rev. Lett. {\bf 81}, 933 (1998).

\bibitem{Strecker} K. E. Strecker, G. B. Partridge, A. G. Truscott, R. G. Hulet, Formation and Propagation of Matter Wave Soliton Trains, Nature (London) {\bf 417}, 150 2002. 

\bibitem{Sardanashvily} G. Sardanashvily, {\it "Noether's Theorems. Applications in Mechanics and Field Theory"}, Springer-Verlag, (2016).

\end{thebibliography}
\end{document}